\documentclass[aps,prl,twocolumn,groupedaddress,showpacs]{revtex4}

\usepackage{amssymb}
\usepackage{graphicx}

\begin{document}

\title{Holographic bound and protein linguistics}

\author{Dirson Jian Li}
\email[]{dirson@mail.xjtu.edu.cn}
\author{Shengli Zhang}

\affiliation{Department of Applied Physics, Xi'an Jiaotong
University, Xi'an 710049, PR China}

\date{\today}

\begin{abstract}

The holographic bound in physics constrains the complexity of life.
The finite storage capability of information in the observable
universe requires the protein linguistics in the evolution of life.
We find that the evolution of genetic code determines the variance
of amino acid frequencies and genomic GC content among species. The
elegant linguistic mechanism is confirmed by the experimental
observations based on all known entire proteomes.

\end{abstract}


\maketitle

The phenomenon of life is governed by the general principles in
physics, so the progress in understanding the physical world may
provide new insight into the origin and evolution of life. The
analogy between biology and linguistics at the level of sequences
hints that the bio-information is processed by underlying linguistic
rules. Several attempts have been made to combine linguistic theory
with biology \cite{ProteinLinguistice}. But the existence of
linguistics in the biomacromolecular sequences needs a physical
explanation. The holographic bound, intimately related to the
holographic principle, came from the deep insights of Bekenstein and
Hawking in 70's \cite{HolographicPrinciple}\cite{Susskind book}. Its
validity is insured by the second law of thermodynamics.
Interestingly, the problem on the existence of linguistics in the
biomacromolecular sequences can be explained by the holographic
bound. In the past decades, the biology has been changed greatly.
Wada advocated ``... to determine the `first principles' of
bio-sciences and link them with the first principles of
non-bio-sciences in order to understand the complex systems.'' and
Gilbert also emphasized the importance of the theoretical methods in
biology \cite{Wada}. Nowadays, the intimacy between biology and
physics is unprecedented. Considering the significant role of
information either in physics or in biology
\cite{Forsdyke}\cite{Infor OriginOfLife}, the gap between physics
and biology may be bridged from the viewpoint of information.

In the post-genomic era, the number of entire proteomes increases
rapidly. We can take all known entire proteomes as samples to study
the global properties of life on our planet. The variance of GC
content \cite{Osawa} is a global property, which varies greatly
among species. We also found another global property of the
evolution of amino acid frequencies though they vary slightly. The
mechanism of the variance of genomic GC content and amino acid
frequencies was a long-standing and far-reaching problem \cite{GC
ratio}. The genetic code evolved in the context of four-letter
alphabet when the $20$ amino acids joined protein sequences
chronologically \cite{EvolCode}. The nature of prime biased AT/GC
pressure and the reason for the correlation of GC content between
total genomic DNA and the 1st, 2nd and 3rd codon positions were
unknown. The profound mechanism behind the variance of amino acid
frequencies has not been studied; it is worse that the amino acid
frequencies are routinely assumed to be constant. All of these basic
problems in biology are solved in our theoretical framework based on
the formal linguistics and the evolution of genetic code.

In this paper, firstly we explain the existence of protein
linguistics and the limited complexity of life in the universe in
terms of the holographic bound. Secondly, a linguistic model is
proposed to reveal the mechanism of the evolution of amino acid
frequencies and genomic GC content as well as the protein length
distribution. The excellent fit between our simulations and the
experimental observations strongly supports the linguistic
mechanism, where the experimental observations are based on the data
of $106$ entire proteomes ($85$ eubacteria, $12$ archaebacteria, $7$
eukaryotes and 2 viruses) in database PEP \cite{PEP} and the data of
GC content in database Genome Properties system \cite{GCcontent}.
The ``information'' is the thread of the paper, which connects
traditionally irrelative problems in physics and biology with each
other.

According to the holographic bound, which states that the
information storage capacity of a spatially finite system is limited
by a quarter of its boundary area measured in Plank area unless the
second law of thermodynamics is untrue, the entropy $S$ in a volume
of radius $R$ satisfies
\begin{equation}
S \leqslant S_{max} \approx (\frac{R}{l_p})^2,
\end{equation}
where $l_p$ is the Plank length. There is much astronomical evidence
that our universe may be headed for an infinite deSitter space. The
holographic bound can be applied to the observable universe with
finite event horizon. Therefore we can estimate the upper limit of
the information storage capability of the observable universe as $
I_{univ} \approx 10^{122} \ bits $ \cite{Susskind book}. In the
point of view in physics, the entropy in our universe is given
primarily by the number of black body cosmic background photons,
$\sim 10^{90}$, which is definitely less than $I_{univ}$.

However, the upper limit of information $I_{univ}$ is not so large
when considering the information of a system of life. Firstly, let
us give a reasonable definition of a ``living'' system from the
viewpoint of information. Many restless functional proteins,
composed of $20$ amino acids (a.a.), distinguish the life from the
lifeless matter. Thus, a general living system $L(n)$ is defined as
a set of all possible proteins with length no more than $n$ a.a.,
each of which is in either the folded state or the unfolded state.
The maximum length $n$ indicates the complexity of the system. Then,
let's calculate the information of the system. The number of states
of $L(n)$ is $ \Omega(n) = \sum_{k=2}^{n}2^{20^{k}}, $ hence its
information is
\begin{equation} I(n) = log_2 \Omega(n)\approx 20^n \ bits.
\end{equation} The exquisite single-chain structure of proteins can
provide much more information storage capacity than lifeless matter.
The upper limit of information $I_{univ}$ forbids $L(n),\ n
> n_0$ to exist in our universe, where $n_0=94$ a.a. such
that $I(n_0) \sim I_{univ}$. Interestingly, the most frequent
protein length for the life on our planet is about $n_0$.

The actual system of life on the planet is not one of $L(n)$,
because the average protein length for different species, ranging
from $250$ a.a. to $500$ a.a., are greater than $n_0$. Let the
actual living system $L_{earth}$ be the set of all possible proteins
on the planet, and suppose $n^*\ (>n_0)$ be the maximum protein
length. $L_{earth}$ must be a proper subset of $L(n^*)$ because the
information of $L_{earth}$ is bounded by $I_{univ}$. According to
the linguistic theory, $L_{earth}$ is a language over the alphabet
of $20$ amino acids. So we have demonstrated the existence of
protein linguistics in terms of the holographic bound that can be
derived from the second law of thermodynamics. In early evolution
when delivering the genetic information from the RNA world to the
DNA-protein world, the holographic bound required the grammars to
allow a very small part of protein sequences and to forbid all the
others. There is a easy way to implement the forbiddance: the
observable universe can not accommodate all the proteins in
$L(n^*)$. In fact, our conclusion is based on the general principle,
freedom of the subtleties of the hierarchy.

If $L(n)$ have an additional property, e.g., chirality, $L^{r}(n)$
($L^{l}(n)$) become the set of proteins composed of right-handed
(left-handed) amino acids. Let ${\mathbf C}(n)$ be the set of all
possible chiral $L(n)$. Only up to $n=2$ a.a., the information of
${\mathbf C}(2)$, i.e., $ I^{C}(2) = \Omega(2) \approx 10^{120} \
bits $ is near $I_{univ}$. Chirality might have brought too much
redundant information; broken symmetry was the solution. So entropy
bound is a strong law which constrains the forms of possible life in
general. Let us imagine the most complex creature with the same
height of us be the one whose genetic information is stored in Plank
scale. The information stored in its body can be estimated as
$1/l_p^3 \sim 10^{105} \ bits$, which violates the holographic
bound. So such complex creature can not exist.

So far, we are aware that the linguistics must play a significant
role in generating proteins in the primordial time. In the
following, a linguistic model is proposed to simulate the variance
of the amino acid frequencies and the genomic GC contents as well as
the protein length distributions.

The model consists of three parts: ({\it i}) generate protein
sequence by tree adjoining grammar \cite{TAG}; ({\it ii}) set amino
acid for the leaves of grammars in (i) according to the tree of
genetic code multiplicity (can be obtained from symmetry analysis,
see \cite{AA Tree}) with consideration of the amino acid chronology
\cite{AA Chronology}; and ({\it iii}) translate the protein
sequences to the DNA sequences according to genetic code chronology
\cite{GeneticCode Chronology}. The evolution of genetic code is the
core of the model. There is a variant {\it t} in the model, which
represents the time in evolution. A proteome for a species is
defined as many a protein generated by the model with fixed {\it t},
so {\it t} also identifies species in the model. Thus, the amino
acid frequencies and the average protein length for a species can be
calculated. The evolutionary trends of the amino acid frequencies
can be determined when proteomes are generated at different time
{\it t}. We can also simulate the evolution of genomic GC content
after translating the protein sequences to DNA sequences.

\begin{figure}
\centering{
\includegraphics[width=86mm]{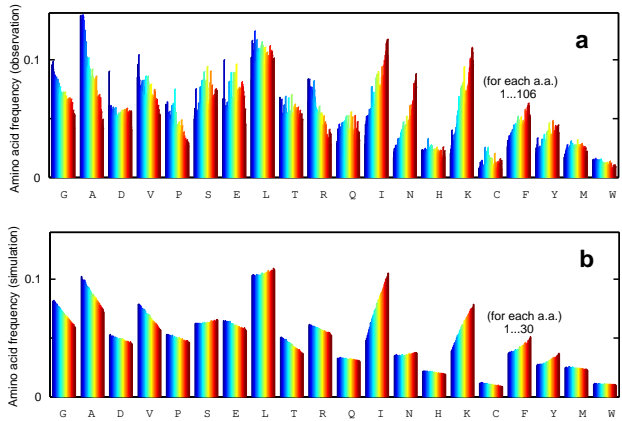}
} \label{fig1} \caption{\small {\bf Evolution of amino acid
frequencies.} The $20$ amino acids are aligned chronologically. The
variance for each amino acid in simulation fits the experimental
observation. ({\bf a}) Experimental observation base on the data of
$106$ species. For each amino acid, the $106$ species are aligned
from left to right by $R_{10/10}$. ({\bf b}) Simulation by the
linguistic model. The $30$ simulated proteomes are aligned by {\it
t}, which increases from $0.02$ to $0.40$ by equal steps. (The
simulations in Fig. 1, Fig. 2 and Fig. 4 are obtains together by the
linguistic model.)}
\end{figure}

The evolution of amino acid frequencies can be explained by the
model. According to the consensus chronology of amino acids to
recruit into the genetic code from the earliest to the latest
\cite{AA Chronology}: G, A, D, V, P, S, E, L, T, R, Q, I, N, H, K,
C, F, Y, M, W, we sort the $106$ species by the ratio $R_{10/10}$ of
average frequency for $10$ later amino acids to average frequency
for $10$ earlier amino acids. Then we obtain the evolutionary trends
of amino acid frequencies: the frequencies of G, A, D, V, P, T, R,
H, W decrease, while the frequencies of S, E, I, N, K, F, Y increase
and the frequencies of L, Q, C, M do not vary obviously (Fig. 1a).
The variance of amino acid frequencies are amazingly monotonic by
and large. Therefore, it is reasonable to assume that a mechanism
underlies the evolution of amino acid frequencies.

The simulation of our linguistic model [Fig. 1b] agrees with the
data of 106 species [Fig. 1a] not only in the evolutionary trends
but also in the variance magnitudes for most of the amino acids.
Note that no parameter is added on purpose in the model to alter the
trend for a certain amino acid. The evolution of amino acid
frequency are sensitive to the amino acid multiplicities \cite{AA
Tree}, any disobedience of which would spoil the results. Therefore,
it is the evolution of genetic code that determines the evolution of
amino acid frequencies. An important property of the model is that
the parameters of amino acid frequencies are constant, which
indicates that the variance of amino acid frequencies developed
during a short period. It agrees that the genetic code had
accomplished quickly. Recently, Jordan et al observed the
contemporary amino acid gain and loss, about which there were
different explanations \cite{Jordan}. We believe that the evolution
of genetic code drives the amino acid gain and loss.

\begin{figure}
\centering{
\includegraphics[width=86mm]{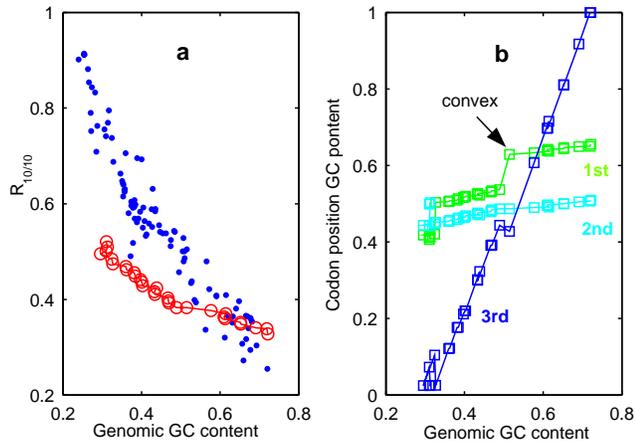}
} \label{fig2} \caption{\small {\bf Evolution of genomic GC
content.} ({\bf a}) Relationship between genomic GC content and
$R_{10/10}$ for the species in database PEP (dots) and its
simulation by the linguistic model (solid line, also decreasing).
({\bf b}) Simulation of the correlation of the GC content between
total genomic DNA and the 1st, 2nd, and 3rd codon positions, which
agrees with the experimental observation in detail (see fig. 5 in
\cite{GC123}, fig. 2 in \cite{Osawa} and fig. 9-1 in
\cite{Forsdyke}).}
\end{figure}

The genomic GC content decreases linearly with $R_{10/10}$ for the
species in database PEP. The simulation of our model agrees
qualitatively with this experimental observation [Fig. 2a]. In our
model, the evolution of amino acid frequency and the evolution of
genomic GC content are driven by a common variant {\it t}. A protein
sequence generated at later time {\it t} corresponds to the DNA
sequence translated using the later codons \cite{GeneticCode
Chronology}, which results in the relationship between genomic GC
content and $R_{10/10}$.

And the simulation of the correlation of GC content between total
genomic DNA and the first, second, and third codon positions [Fig.
2b] also agrees with the results based on the data of completed
genomes \cite{Osawa}\cite{Forsdyke}\cite{GC123}, where the
correlation slope of the third codon position is much greater than
that of the first and the second positions. There is a
characteristic convex in the middle of the line of the simulated GC
content for the first codon position, which agrees dramatically with
the experimental observations
\cite{Osawa}\cite{Forsdyke}\cite{GC123}. In the table of codon
chronology \cite{GeneticCode Chronology}, G and C (A and U) occupy
all the third positions of earliest (latest) codons for 20 amino
acids, while the bases appear about equally for the first and second
positions. Therefore, the correlation slope for the first and second
positions vary slightly while the slope for the third position
varies greatly. And the lower limit $l \sim 0.3$ and upper limit $u
\sim 0.7$ of the GC content among species can be explained
similarly; $l+u=1$ is required in theory.

\begin{figure}
\centering{
\includegraphics[width=86mm]{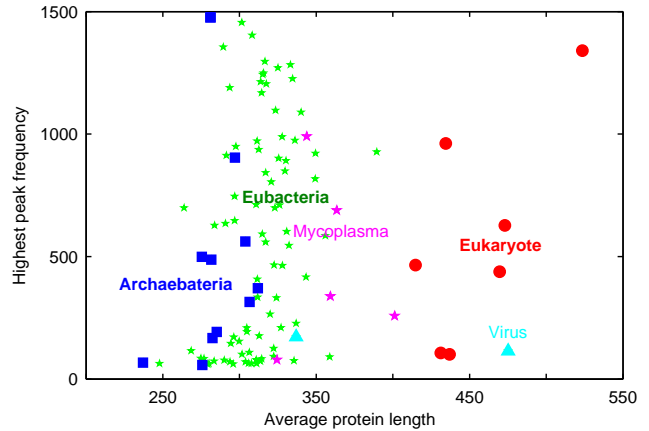}
} \label{fig3} \caption{\small {\bf Rainbow distribution.}
Relationship between the average protein length and the highest
frequency of the discrete fourier transformation of protein length
distribution (the cutoff is $3000$) for each of the $106$ species.
The distribution of the species from three domains likes a rainbow.
Even for the group of closely related species such as mycoplasmas
(belonging to eubacteria), their distribution also form an ``arch''
of the rainbow.}
\end{figure}

The linguistic mechanism can also be supported by the distribution
of protein length. When observing the distribution of species in the
space of average protein length and the highest frequency of
discrete fourier transformation of protein length distribution, we
unexpectedly found that the species for the three domains gathered
in three parallel arches respectively, which likes a rainbow [Fig
3]. This indicates that the fluctuations in protein length
distributions can not be products of a stochastic process. The
characters of the protein length distribution (bell-shape profile,
periodic-like fluctuations) have been simulated by the linguistic
model, which should be intrinsic properties related to underlying
grammars \cite{to appear}.

\begin{figure}
\centering{
\includegraphics[width=86mm]{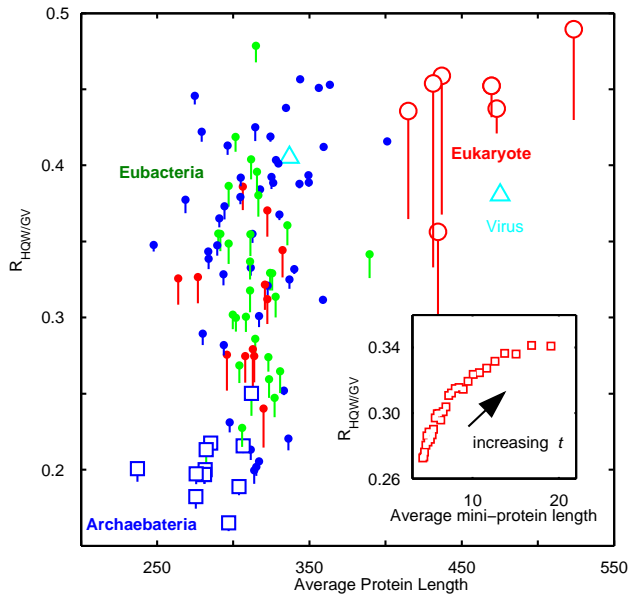}
} \label{fig4} \caption{\small {\bf The evolutionary flow.}
Relationship between average protein lengths and $R_{HQW/GV}$ for
the $106$ species. The species of three domains (Archaebacteria:
blue square, Eubacteria: dot, Eukaryotes: red circle) gather
together in respective regions and all the species form an
evolutionary flow. The proteome size is represented proportionally
by the tail under each species (big: red, middle: green and small:
blue); species with big genome sizes locate in the midstream of the
evolutionary flow. ({\bf Embedded}) Simulation of the evolutionary
flow, whose (upward) bending direction agrees with the direction of
the experimental observation.}
\end{figure}

We also find the relationship between the average protein length and
the ratio of amino acid frequencies. The species of three domains
gather in different regions in the space of the average protein
length and the ratio $R_{HQW/GV}$ of average frequency for several
later amino acids (H, Q, W) to average frequency for several earlier
ones (G, V) [Fig. 4]. The points of all species form a bending line
[Fig. 4], which can be explained as an evolutionary flow in that
({\it i}) the species with large (small) genome locate in the
midstream (margin) of the flow [Fig. 4] and ({\it ii}) the
(rightward) evolutionary direction parallels the directions of
decreasing correlations of protein length distributions among groups
of the closely related species. The evolutionary flow can be
simulated by our model [Fig. 4, Embedded]. The evolutionary
direction and the bending direction in the simulation agree with the
evolutionary flow of the $106$ species.

In conclusion, the holographic bound improves our understanding of
life, which supervises the maximum complexity of life. Linguistics
is necessary in storage of information in the protein/DNA sequences.
We show that the particular variance of amino acid frequencies and
GC content for the contemporary species are the products of certain
genetic code multiplicity and amino acid chronology evolved in
primordial time. The linguistic model succeeds not only in the
simulations of respective aspects (amino acid frequencies [Fig. 1],
GC content [Fig. 2b], protein length distribution [Fig. 3]) but also
in their relationships (amino acid frequencies and GC content [Fig.
2a], amino acid frequencies and protein length distribution [Fig.
4]). So the thorough and detailed fit between simulations and
experimental observations confirms the validity of the linguistic
framework, which is grounded in general principles in physics.

\begin{acknowledgments}

We thank Hefeng Wang, Liu Zhao, Donald R. Forsdyke, Zhenwei Yao for
valuable discussions. Supported by NSF of China Grant No. of
10374075.

\end{acknowledgments}

\end{document}